\documentclass[11pt]{article}
\usepackage{moriond,epsfig}

\def\farcs{\hbox{$.\!\!^{\prime\prime}$}}
\def\fs{\hbox{$.\!\!^{\rm s}$}}

\begin{document}
\vspace*{4cm}
\title{RECENT ARECIBO TIMING OF THE \\ RELATIVISTIC BINARY PSR~B1534+12}

\author{ I. H. STAIRS }

\address{University of Manchester, Nuffield Radio Astronomy Laboratories,\\
Jodrell Bank, Macclesfield, Cheshire, SK11 9DL, UK}

\author{ D. J. NICE, S. E. THORSETT, J. H. TAYLOR }

\address{Joseph Henry Laboratories and Physics Department,  Princeton University, \\ Princeton, NJ 08544 USA}

\maketitle\abstracts{
We present a new timing solution for PSR~B1534$+$12, based on
coherently-dedispersed observations at Jodrell Bank and, recently,
Arecibo.  The new data have resulted in improved measurements of the
post-Keplerian timing parameters, including the orbital period
derivative, $\dot P_b$.  At present, the poorly-known distance to the
pulsar limits the precision of the measurement of the intrinsic $\dot
P_b$, and hence the strength of the test of general relativity that
results from this binary system.  By assuming that general relativity
is the correct theory of gravity, we may invert the test and find an
improved value of the pulsar distance. }

\section{Introduction}\label{sec:intro}

Double-neutron-star binary pulsars in close, highly eccentric orbits
have long provided the best strong-field tests of the predictions of
gravitational theories.  The timing analysis of pulsar signals permits
the measurement of five Keplerian orbital elements as well as a number
of post-Keplerian (PK) orbital parameters.  The PK parameters can be
determined using a theory-independent procedure in which the masses of
the two stars are the only unknowns.\cite{dt92} Each of the PK
parameters depends on the masses in a different way; consequently, if
any two of them are measured, the relevant parameters of the two-body
system are fully determined within any gravitational theory.  If three
or more PK parameters can be measured, a test of the gravitational
theory results from the overdetermined system.  

For the first binary pulsar, PSR~B1913+16, the PK parameters
$\dot\omega$ (rate of advance of periastron), $\gamma$ (time dilation
and gravitational redshift) and $\dot{P_b}$ (orbital period
derivative) have been measured and found to be in excellent agreement
with the predictions of general
relativity.\cite{tw89,dt91,tay94,tay99} PSR~B1534+12, discovered in
1990,\cite{wol91a} is a comparable system, with an eccentric 10.1-hour
orbit.  PSR~B1534+12 is significantly brighter than PSR~B1913+16, and
its pulse profile has a narrow peak, permitting timing measurements of
nearly five times greater accuracy.  Because the orbit is nearly
edge-on as viewed from the Earth, the PK parameters $r$ and $s$ (the
``range'' and ``shape'' of the Shapiro time delay) are measurable, in
addition to $\dot\omega$, $\gamma$ and $\dot P_b$.  The resulting
overdetermination of the orbit leads to a non-radiative test of
gravitation theory in the strong-field regime, complementing the
$\dot\omega$-$\gamma$-$\dot P_b$ test for PSR~B1913+16.\cite{twdw92}

Previously published timing measurements of PSR~B1534+12 were made
with the Arecibo 305\,m telescope through early 1994,\cite{arz95} when
the telescope went out of normal service for a major upgrading; and
from 1994 through 1997 with the 43\,m telescope at Green Bank, West
Virginia and the 76\,m Lovell telescope at Jodrell Bank,
England.\cite{sac+98} With the recent reopening of Arecibo, we have
been able to conduct a new series of observations of this pulsar in
1998 July, and improve on our previously reported timing results; in
particular, we have reduced the uncertainty on the observed value of
the PK parameter $\dot P_b$ by 35\%, to approximately 5\% of the
expected GR value.

\section{Observations}\label{sec:obs}

A complete discussion of the observing procedures and instrumentation
may be found in ref.~9.  We only note here that over the years, new
instrumentation was developed to remove more perfectly the
pulse-smearing effects of interstellar dispersion.  While the
1990-1994 Arecibo observations and the Green Bank observations used
filterbanks to divide the observing bandpass into small channels, the
later observations were carried out with an improved ``coherent
dedispersion'' system, which sampled the raw telescope voltages and
then removed the effects of dispersion using a software filter.  The
more recent observations therefore yield significantly better timing
precision.

All observations involved folding the pulse signal over several
minutes at the predicted topocentric pulse period to produce a
total-intensity profile.  Cross-correlation with a standard template,
using a least-squares method in the Fourier transform domain, yielded
a time offset.\cite{tay92} This offset was added to the time of the
first sample of a period near the middle of the integration, thereby
yielding an effective pulse arrival time.  A different standard
template was used for each observing system and frequency; they were
made by averaging the available profiles over several hours or more.
Uncertainties in the TOAs were estimated from the least squares
procedure, and also from the observed scatter of the TOAs within 30
minutes of each one.  Each observatory's local time standard was
corrected retroactively to the UTC timescale, using data from the
Global Positioning System (GPS) satellites.

\section{The Timing Model}\label{sec:timing}

A pulse received on Earth at topocentric time $t$ is emitted at a time
in the comoving pulsar frame given by
\begin{equation}
T =  t-t_0+\Delta_C-D/f^2 + \Delta_{R\odot} + \Delta_{E\odot}
  -\Delta_{S\odot} - \Delta_R - \Delta_E - \Delta_S\,.
\label{eqn:orbit}
\end{equation}
Here $t_0$ is a reference epoch and $\Delta_C$ is the offset between
the observatory master clock and the reference standard of terrestrial
time.  The dispersive delay is $D/f^2$, where $D={\rm
DM}/2.41\times10^{-4}$, with dispersion measure DM in cm$^{-3}$pc,
radio frequency $f$ in MHz, and the delay in seconds.  Finally,
$\Delta_{R\odot}$, $\Delta_{E\odot}$, and $\Delta_{S\odot}$ are
propagation delays and relativistic time adjustments for effects
within the solar system, and $\Delta_R$, $\Delta_E$ and $\Delta_S$ are
similar terms accounting for phenomena within the pulsar's
orbit.\cite{dt92,tw89,dd86}   The orbital
$\Delta$ terms are defined by:
\begin{eqnarray}
\Delta_R & = & x \sin\omega (\cos u -e)  + x ( 1-e^2)^{1/2}
       \cos\omega \sin u, \\
\Delta_E & = & \gamma \sin u, \\
\Delta_S & = & -2r \ln \left\{ 1-e\cos u - s \left[
  \sin\omega (\cos u - e) + (1-e^2)^{1/2} \cos\omega \sin u
  \right] \right\}\,. 
\end{eqnarray}
These are written in terms of the
eccentric anomaly $u$ and true anomaly $A_e(u)$, and the time
dependence of $\omega$, which are related by:
\begin{eqnarray}
u-e\sin u & = & 2\pi \left[ \left( {{T-T_0}\over{P_b}} \right) -
  {{\dot P_b}\over 2} \left( {{T-T_0}\over{P_b}} \right)^2
  \right], \\
A_e(u) & = & 2 \arctan \left[ \left( {{1+e}\over{1-e}}
  \right)^{1/2} \tan {u\over2} \right], \\
\omega & = & \omega_0 + \left(\frac{P_b\,\dot{\omega}}{2\pi}\right) 
         A_e(u).  
\end{eqnarray}
At a given time $t$, then, the propagation delay across the pulsar
orbit is calculated by a model which incorporates ten parameters
implicitly defined in the above equations: five Keplerian parameters
($x$, $\omega$, $T_0$, $P_b$, $e$) and five PK parameters
($\dot{\omega}$, $\dot{P_b}$, $\gamma$, $r$, $s$).  These quantities,
in conjunction with a simple time polynomial to model the spin of the
pulsar and with astrometric parameters to model the propagation of the
signal across the solar system, constitute the free parameters to be
fit in the theory-independent timing model.

In a particular theory of gravity, the five PK parameters can be
written as functions of the pulsar and companion star masses, $m_1$
and $m_2$, and the well-determined Keplerian parameters. In general
relativity, the equations are:\cite{dt92,tw89,dd86}
\begin{eqnarray}
\dot\omega &=& 3 \left(\frac{P_b}{2\pi}\right)^{-5/3}
  (T_\odot M)^{2/3}\,(1-e^2)^{-1}\,, \label{eq:omdot} \\
\gamma &=& e \left(\frac{P_b}{2\pi}\right)^{1/3}
  T_\odot^{2/3}\,M^{-4/3}\,m_2\,(m_1+2m_2) \,, \\
\dot P_b &=& -\,\frac{192\pi}{5}
  \left(\frac{P_b}{2\pi}\right)^{-5/3}
  \left(1 + \frac{73}{24} e^2 + \frac{37}{96} e^4 \right)
  (1-e^2)^{-7/2}\,T_\odot^{5/3}\, m_1\, m_2\, M^{-1/3}\,,
  \label{eq:pbdot} \\
r &=& T_\odot\, m_2\,, \\
s &=& x \left(\frac{P_b}{2\pi}\right)^{-2/3}
  T_\odot^{-1/3}\,M^{2/3}\,m_2^{-1}\,. \label{eq:s}
\end{eqnarray}
Here the masses $m_1$, $m_2$, and $M\equiv m_1+m_2$ are expressed in
solar units, and we use the additional shorthand notations
$s\equiv\sin i$ and $T_\odot\equiv GM_\odot/c^3 = 4.925490947\,\mu$s,
where $i$ is the angle between the orbital angular momentum and the
line of sight, $G$ the Newtonian constant of gravity, and $c$ the
speed of light.

\section{Arrival Time Analysis}

We used the standard {\sc tempo} analysis software\,\cite{tw89} together
with the JPL~DE200 solar-system ephemeris\,\cite{sta90} to fit the
measured pulse arrival times to the timing model with a least-squares
technique.  Arbitrary offsets were allowed between the different data
sets to allow for frequency-dependent changes in pulse shape and
slight differences in the standard profile alignments.  To partially
compensate for small systematic errors, uncertainties in TOAs were
increased (in quadrature) by 2.9, 17.1, 20.5, 7.0, and 6.0\,$\mu$s,
respectively, when calculating weights for the 1990-94 Arecibo,
Jodrell Bank, Green Bank data, 1998 Arecibo 430\,MHz and 1998 Arecibo
1420\,MHz sets.

Results for the astrometric, spin, and dispersion parameters of PSR
B1534+12 are presented in Table~\ref{tab:astspin}.  We fit the data to
two models of the pulsar orbit.  The theory-independent ``DD''
model\,\cite{dd86} treats all five PK parameters defined in
\S\ref{sec:timing} as free parameters in the fit.  Alternatively, the
``DDGR'' model\,\cite{tw89,tay87a} assumes general relativity to be
correct and uses equations~\ref{eq:omdot} through~\ref{eq:s} to link
the PK parameters to $M\equiv m_1+m_2$ and $m_2$; consequently it
requires only two post-Keplerian free parameters.

\begin{table}[t]
\caption{Astrometric, spin, and dispersion
parameters for PSR B1534+12.  Figures in parentheses are 
uncertainties in the last digits
quoted, and italic numbers represent derived quantities.\label{tab:astspin}}
\vspace{0.4cm}
\begin{center}
\begin{tabular}{|l|l|}
\hline
Right ascension, $\alpha$ (J2000)  & 
  $15^{\rm h}\,37^{\rm m}\,09\fs959952(16)$ \\
Declination, $\delta$ (J2000)    & 
  $11^\circ\,55'\,55\farcs6561(3)$ \\
Proper motion in R.A., $\mu_\alpha$ (mas\,yr$^{-1}$)    & 
1.5(2) \\
Proper motion in Dec., $\mu_\delta$ (mas\,yr$^{-1}$)    & 
$-$25.6(3) \\
Parallax, $\pi$ (mas)  & $<1.5$ \\
 &  \\
Pulse period, $P$ (ms)    & 
37.9044404878550(14)\\
Period derivative, $\dot P$ $(10^{-18})$  & 2.42250(3) \\
Epoch (MJD)  & 48778.0 \\
 &  \\
Dispersion measure, DM (cm$^{-3}$pc)    & 
11.6152(16) \\
DM derivative (cm$^{-3}\mbox{pc\,yr}^{-1}$)    & 
$-$0.00004(53) \\
 &  \\
Galactic longitude $l$ (deg)   & {\it 20.0} \\ 
Galactic latitude $b$ (deg)   & {\it 47.8}  \\
Composite proper motion, $\mu$ (mas\,yr$^{-1}$) & {\it 25.6(3)} \\
Galactic position angle of $\mu$ (deg)   & {\it 238.7(4)}  \\
\hline
\end{tabular}
\end{center}
\end{table}

\begin{table}[t]
\caption{Orbital parameters of PSR B1534+12 in the DD and DDGR
models. Figures in parentheses are
uncertainties in the last digits quoted, and italic numbers represent
derived quantities. \label{tab:orbparms}}
\vspace{0.4cm}
\begin{center}
\begin{tabular}{|l|l|l|}
\hline
& DD model & DDGR model \\
\hline
Orbital period, $P_b$ (d)   & 0.42073729933(3) & 0.42073729933(2) \\ 
Projected semi-major axis, $x$ (s)   & 3.729464(3) & 3.7294638(5) \\
Eccentricity, $e$   & 0.2736775(5) & 0.2736774(2) \\ 
Longitude of periastron, $\omega$ (deg)   & 267.44738(10) 
  & 267.44744(9) \\
Epoch of periastron, $T_0$ (MJD)   & 48777.82595097(6) 
  & 48777.82595096(6) \\
 & & \\
Advance of periastron, $\dot\omega$ (deg\,yr$^{-1}$)   
  & 1.755794(19) & {\it 1.75580} \\
Gravitational redshift, $\gamma$ (ms)   & 2.071(6) & {\it 2.067} \\
Orbital period derivative, $(\dot P_b)^{\rm obs}$ $(10^{-12})$  &
  $-$0.131(9) & {\it $-$0.1924} \\
Shape of Shapiro delay, $s$   & 0.983(8) & {\it 0.9762} \\
Range of Shapiro delay, $r$ ($\mu{\rm s}$)   & 6.3(1.3) & 
 {\it 6.6} \\
 & & \\
Total mass, $M=m_1+m_2$ ($M_{\odot}$)   &  & 2.67845(4) \\
Companion mass, $m_2$ ($M_{\odot}$)     &  & 1.344(2) \\
Excess $\dot P_b$ $(10^{-12})$          &  & 0.061(9) \\
\hline
\end{tabular}
\end{center}
\end{table}

Table~\ref{tab:orbparms} presents our adopted orbital parameters.
Uncertainties given in the table are approximately twice the formal
``$1\,\sigma$'' errors from the fit; we believe them to be
conservative estimates of the true 68\%-confidence uncertainties,
including both random and systematic effects.  The Keplerian orbital
parameters include the period $P_{b}$, projected semi-major axis
$x\equiv a_1\sin i/c$, eccentricity $e$, longitude of periastron
$\omega$, and time of periastron $T_0$.  These quantities are followed
by the measured post-Keplerian parameters relevant to each of the two
models.  The DDGR solution includes a parameter called ``excess $\dot
P_b$,'' which accounts for an otherwise unmodeled acceleration
resulting from galactic kinematics.

The best estimates of the masses of the pulsar and its companion come
from the DDGR solution.  We find the masses to be
$m_1=1.344\pm0.002~M_\odot$ and $m_2=1.335\pm0.002~M_\odot$.  For the
sake of comparison Table~\ref{tab:orbparms} lists, in italic numbers,
computed PK parameter values corresponding to the measured masses in
the DDGR fit.  The fitted and derived parameter values are in
accord, indicating good agreement of the theory-independent
solution with general relativity.

\begin{figure}[t]
\begin{center}
\psfig{figure=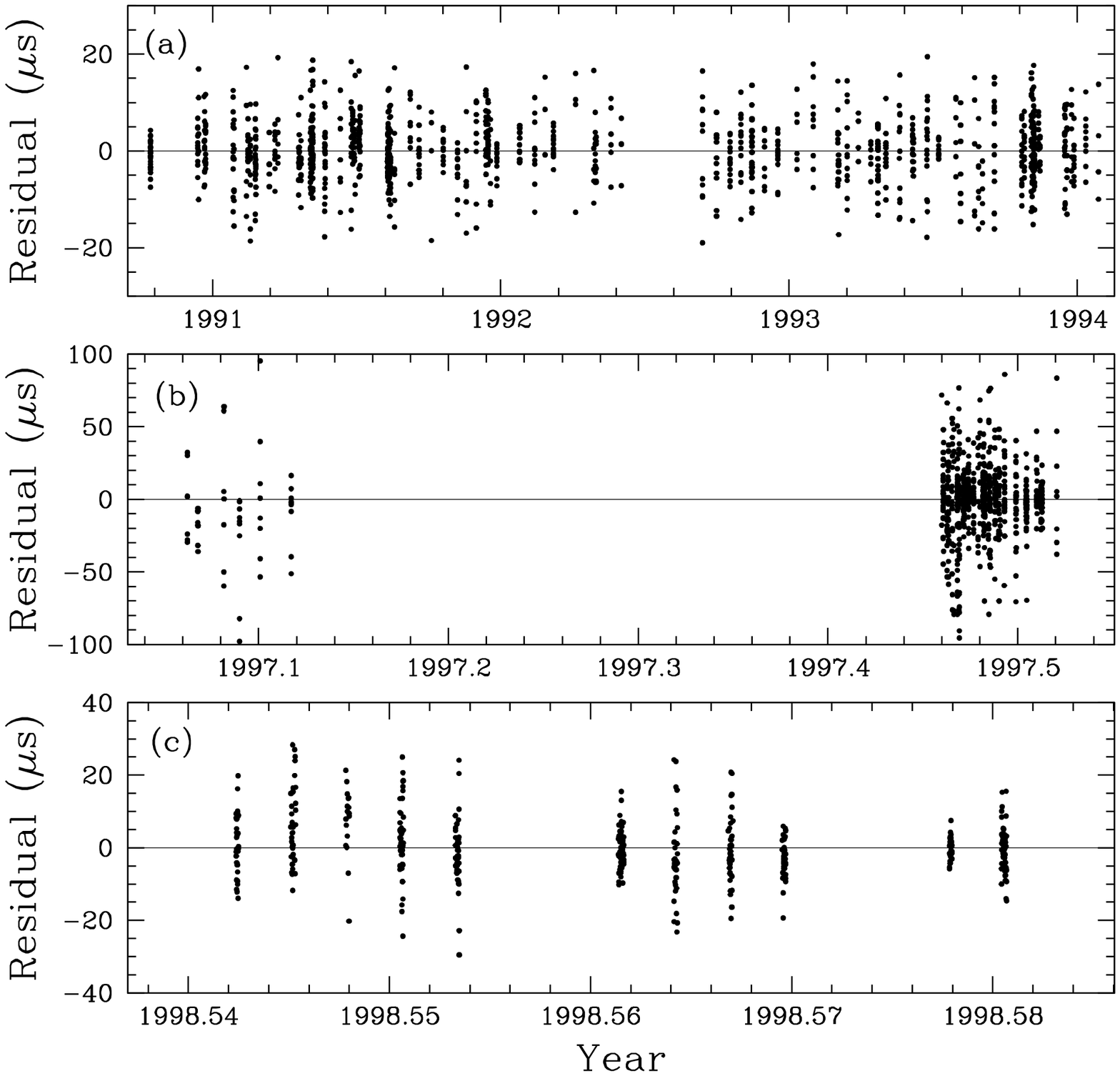,height=4.5in}
\end{center}
\caption{Post-fit residuals versus date for 
(a) Arecibo 1400\,MHz, (b) Jodrell Bank and (c) Arecibo 430\,MHz data.
\label{fig:dayres}}
\end{figure}

Figure~\ref{fig:dayres} shows the post-fit residuals for the
1990--1994 Arecibo 1400\,MHz data, the Jodrell Bank data, and the
recent Arecibo 430\,MHz data, plotted as functions of date.  Even
within a single data set, the TOA uncertainties can vary by factors of
three or more because of scintillation-induced intensity variations.
We have not attempted to show these differences in data quality by
means of error bars in the residual plots.

\section{Discussion}\label{sec:disc}

Before the observed $\dot P_b$ can be compared to the theoretical
value, we must apply a correction which accounts for the relative
acceleration of the center-of-mass frame of the binary pulsar system
and the solar system barycenter.  An expression for the most
significant bias, which arises from galactic kinematic effects, is
derived in ref.~3.  This bias can be written as the sum of terms
arising from acceleration toward the plane of the Galaxy, acceleration
within the plane of the Galaxy, and an apparent acceleration due to
the proper motion of the binary system:
\begin{equation}\label{eqn:gal}
\left(\frac{\dot{P_b}}{P_b}\right)^{\rm gal} = -\,\frac{a_z\sin b}{c}
 \,-\,\frac{v_0^2}{cR_0} \left[\cos l + 
  \frac{\beta}{\sin^2 l + \beta^2}\right] +\mu^2\frac{d}{c}.
\end{equation}
Here $a_z$ is the vertical component of galactic acceleration, $l$ and
$b$ the galactic coordinates of the pulsar, $R_0$ and $v_0$ the Sun's
galactocentric distance and galactic circular velocity, $\mu$ and $d$
the proper motion and distance of the pulsar; we use the short-hand
notation $\beta=d/R_0 - \cos l$.  The pulsar distance can be estimated
from the dispersion measure, together with a smoothed-out model of the
free electron distribution in the Galaxy.\cite{tc93} This model yields
$d\approx0.7$\,kpc for PSR B1534+12, with an uncertainty of perhaps
0.2\,kpc.  At this distance we estimate
$a_z/c=(1.60\pm0.13)\times10^{-19}\,\mbox{s}^{-1}$.\cite{kg89}
Following ref.~3, we assume a flat galactic
rotation curve and take $v_0=v_1=222\pm20\,$km\,s$^{-1}$ and $R_0 =
7.7\pm0.7$\,kpc.  Then, summing the terms in equation~(\ref{eqn:gal})
and multiplying by $P_b$, we find the total kinematic correction to be
\begin{equation}
\left(\dot P_b\right)^{\rm gal} = (0.038\pm0.012)\times10^{-12}\,.
\end{equation}
The uncertainty in this correction is dominated by the uncertainty in
distance, which is only roughly estimated by the dispersion-measure
model.

\begin{figure}[t]
\begin{center}
\psfig{figure=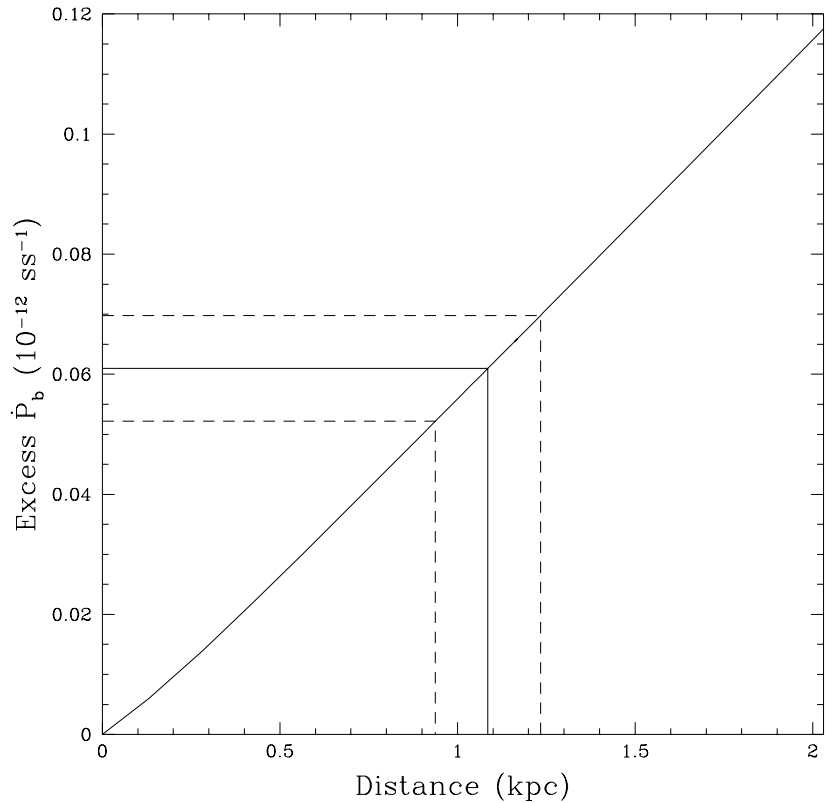,height=4.5in}
\end{center}
\caption{Relation between measured ``excess $\dot{P_b}$'' and pulsar 
distance for the PSR~B1534+12 system.  Dashed lines indicate the 68\% 
confidence ranges of the measured values.\label{fig:pbdotd}}
\end{figure}

Our measurement of the intrinsic rate of orbital period decay is
therefore
\begin{equation}
\left(\dot P_b\right)^{\rm obs} - \left(\dot P_b\right)^{\rm gal}
        = (-0.169\pm0.015)\times 10^{-12}\,.
\end{equation}
Under general relativity, the orbital period decay due to
gravitational radiation damping, $(\dot{P_b})^{\rm GR}$, can be
predicted from the masses $m_1$ and $m_2$ (eq.~\ref{eq:pbdot}), which
in turn can be deduced from the high precision measurements of
$\dot{\omega}$ and $\gamma$.  The expected value is
\begin{equation}
\left(\dot P_b\right)^{\rm GR} = -0.192\times 10^{-12}\,.
\end{equation}
Although the measured value in the pulsar center-of-mass frame differs
from this prediction by 1.5 standard deviations, it can be brought
into good agreement by increasing the pulsar distance to slightly over
1\,kpc.  Stated another way, we can assume that GR is the correct
theory of gravity, measure the ``excess $\dot P_b$'' for the system as
described above and presented in Table~\ref{tab:orbparms}, and then
invert equation~\ref{eqn:gal} to determine the pulsar
distance.\cite{bb96} Figure~\ref{fig:pbdotd} shows the relation
between the two quantities; using this approach we obtain
$d=1.08\pm0.15$~kpc (68\% confidence limit).  The uncertainty is
dominated by the measurement uncertainty of $(\dot P_b)^{\rm obs}$,
rather than uncertainties in the galactic rotation parameters or the
acceleration $a_z$.  Continued timing of this system should lead to a
much more precise distance measurement; in fact, the recent Arecibo
observations have already reduced the uncertainty by 35\% over our
previously published value.\cite{sac+98}

We note that the timing solution provides a second, independent
constraint on the distance.  The upper limit on parallax,
$\pi<1.5$\,mas (Table~\ref{tab:astspin}) constrains the distance to
$d>0.67$\,kpc.  While the parallax distance has less precision than
the kinematic distance, it is reassuring that these measurements are
in agreement.

An accurate distance for PSR~B1534+12 is of considerable interest
because this system, along with PSR~B1913+16, is of prime importance
in estimating the rate of coalescence of binary neutron-star pairs in
a typical galaxy.  Early estimates of the inspiral rate used much
smaller distances for PSR~B1534+12, including 0.4~kpc,\cite{nps91}
0.5~kpc,\cite{phi91} and 0.7~kpc,\cite{cl95,vl96} leading to a low
estimate of its intrinsic luminosity.  Our distance, compared to that
estimated from the dispersion measure, increases the volume per
B1534+12-type object by a factor of 2.5 to 4, depending on the unknown
scale height of such systems in the Galaxy; this has helped lead to
improved estimates of the neutron-star--neutron-star coalescence
rate.\cite{sac+98,acw99,pri99,kal99}

\begin{figure}[t]
\begin{center}
\psfig{figure=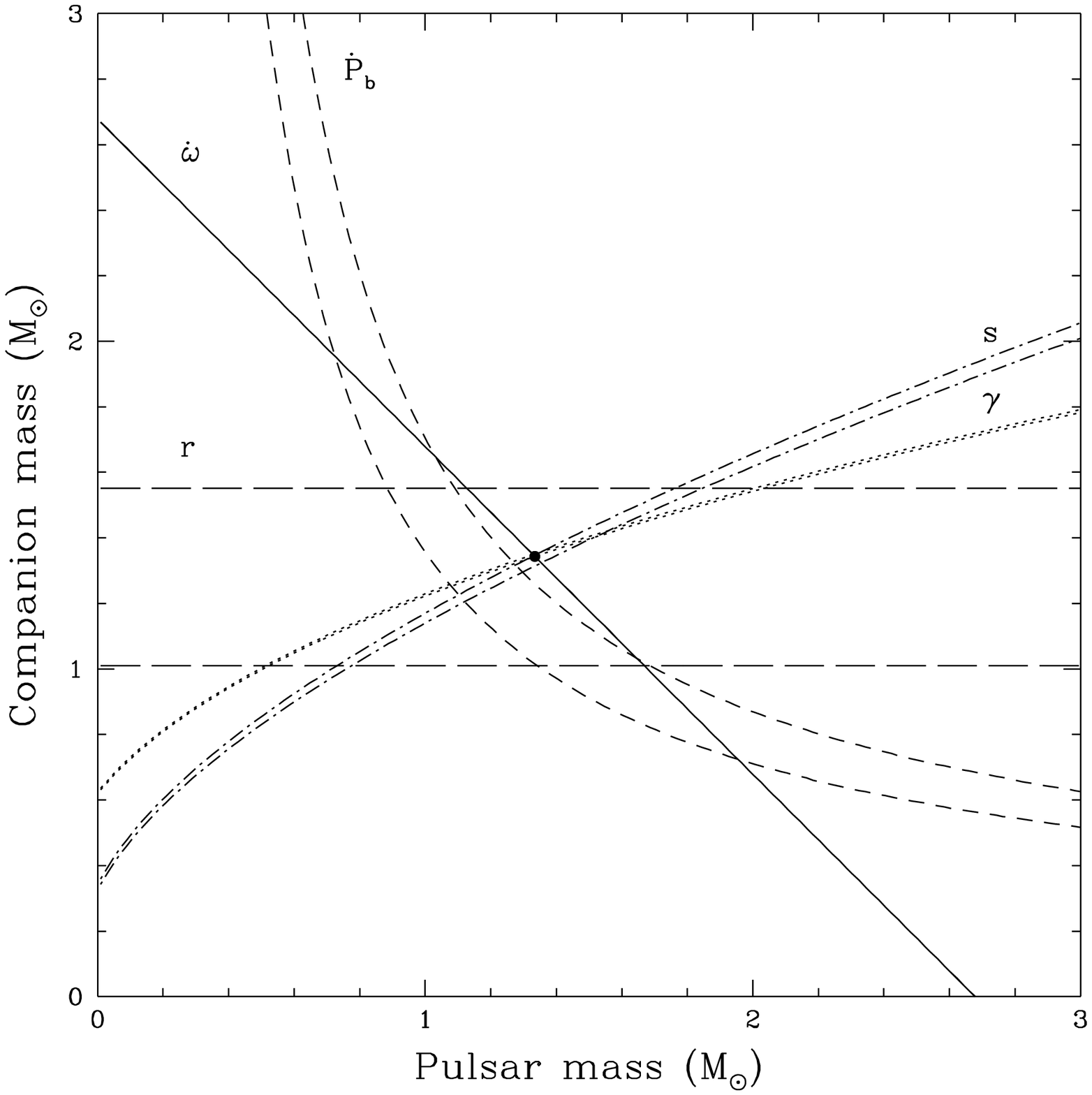,height=4.5in}
\end{center}
\caption{Mass-mass diagram for the PSR~B1534+12
system.  Labeled curves illustrate 68\% confidence ranges of the DD
parameters listed in Table~\ref{tab:orbparms}.  The filled circle
denotes the component masses according to the DDGR solution.  A
kinematic correction for assumed distance $d=0.7\pm0.2\,$kpc has been
subtracted from the observed value of $\dot{P_b}$.  A slightly larger
distance removes the small apparent discrepancy. \label{fig:massmass}}
\end{figure}

\subsection{Test of Relativity}

This pulsar provides the second test of general relativity based on
the $\dot\omega$, $\gamma$, and $\dot P_b$ parameters of a binary
pulsar system, and is the first double-neutron-star binary to permit
significant measurements of the Shapiro-delay parameters $r$ and $s$.
The left-hand sides of equations~(\ref{eq:omdot}--\ref{eq:s})
represent measured quantities, as specified for this experiment in the
``DD'' column of Table~\ref{tab:orbparms}.  If GR is consistent with
the measurements and there are no significant unmodeled effects, we
should expect the five curves corresponding to
equations~(\ref{eq:omdot}--\ref{eq:s}) to intersect at a single point
in the $m_1$-$m_2$ plane.  A graphical summary of the situation for
PSR B1534+12 is presented in Figure~\ref{fig:massmass}, in which a
pair of lines delimit the 68\% confidence limit for each PK parameter
(a single line for $\dot\omega$, whose uncertainty is too small to
show).  A filled circle at $m_1=1.335~M_\odot$, $m_2=1.344~M_\odot$
marks the DDGR solution of Table~\ref{tab:orbparms}, and its location
on the $\dot\omega$ line agrees well with the measured DD values of
$\gamma$ and $s$.  These three parameters therefore provide a unique
test of the purely quasi-static regime of general relativity.  We have
already noted that the DD value of $\dot P_b$ is slightly too small
when corrected to the dispersion-estimated distance.  However, as
discussed above, this discrepancy can be removed by invoking a larger
distance to the pulsar.  Finally, the value of $r$, although presently
little better than a 20\% measurement, also agrees with its expected
value.  This pulsar thus provides a convincing second demonstration of
the existence of gravitational radiation; furthermore, all its timing
parameters are in good accord with general relativity theory.

\section*{Acknowledgments}

We thank Alex Wolszczan, Zaven Arzoumanian, Andrew Lyne and Fernando
Camilo for their extensive contributions to earlier phases of this
project, including the 1990-94 observations at Arecibo, and the
observations at Green Bank and Jodrell Bank.  We also thank
Christopher Scaffidi and Eric Splaver for assistance with recent data
acquisition.  The Arecibo Observatory, a facility of the National
Astronomy and Ionosphere Center, is operated by Cornell University
under a cooperative agreement with the National Science Foundation.
The National Radio Astronomy Observatory is operated by Associated
Universities, Inc., under a cooperative agreement with the US National
Science Foundation.  I.~H.~S. received support from NSERC 1967 and
postdoctoral fellowships.  S.~E.~T. is an Alfred P.~Sloan Foundation
Research Fellow.

\section*{References}

\end{document}